\documentclass[preprint,nofootinbib,aps,superscriptaddress,floatfix]{revtex4}

\usepackage[dvips]{graphics,color}
\begin{document}
\def\LQCD{\Lambda_{\rm QCD}}
\def\lqcd{\Lambda_{\rm QCD}}
\def\xslash#1{{\rlap{$#1$}/}}
\newcommand{\orderalpha}{ {\cal{O}}(\alpha_s)}
\newcommand{\orderalphasqr}{ {\cal{O}}(\alpha_s^2)}
\newcommand{\btoc}{\bar{B} \rightarrow \rm X_c \, \ell \, \bar{\nu}}

\def\ctp#1#2#3{\CTP{\bf #1} (#2) #3}
\def\jetpl#1#2#3{\JETPL{\bf #1} (#2) #3}
\def\nc#1#2#3{\NC{\bf #1} (#2) #3}
\def\np#1#2#3{\NP{\bf B#1} (#2) #3}
\def\pl#1#2#3{\PL B {\bf #1} (#2) #3}
\def\prl#1#2#3{\PRL{\bf #1} (#2) #3}
\def\prd#1#2#3{\PR D {\bf #1} (#2) #3}
\def\prep#1#2#3{\PRep{\bf #1} (#2) #3}
\def\physrev#1#2#3{\PR{\bf #1} (#2) #3}
\def\sjnp#1#2#3{\SJNP{\bf #1} (#2) #3}
\def\nuvc#1#2#3{\NC{\bf #1A} (#2) #3}
\def\blankref#1#2#3{   {\bf #1} (#2) #3}
\def\ibid#1#2#3{{\it ibid,\/}  {\bf #1} (#2) #3}
\def\AP{{\it Ann.\ Phys.\ }}
\def\CMP{{\it Comm.\ Math.\ Phys.\ }}
\def\CTP{{\it Comm.\ Theor.\ Phys.\ }}
\def\IJMP{{\it Int.\ Jour.\ Mod.\ Phys.\ }}
\def\JETPL{{JETP Lett.\ }}
\def\NC{{\it Nuovo Cimento\ }}
\def\NP{{Nucl.\ Phys.\ }}
\def\PL{{Phys.\ Lett.\ }}
\def\PR{{Phys.\ Rev.\ }}
\def\PRep{{Phys.\ Rep.\ }}
\def\PRL{{Phys.\ Rev.\ Lett.\ }}

\newcommand{\nn}{\nonumber}

\title{Jets in Effective Theory: \\Summing Phase Space Logs.}

\author{Michael Trott }\email{mrtrott@physics.ucsd.edu}

\affiliation{Department of Physics, University of California at San Diego,\\[-5pt]
  La Jolla, CA, 92093}

\begin{abstract}
We demonstrate how to resum 
phase space logarithms in the Sterman-Weinberg (SW)
dijet decay rate within the context of Soft Collinear Effective theory (SCET).
An operator basis corresponding to two and three jet events is defined 
in SCET and renormalized. We obtain the RGE of the two and three jet
operators and run the operators from the scale $\mu^2 = Q^2$ to the phase space 
scale $ \mu^2_\delta =  \delta^2 \, Q^2$. This phase space scale, where $\delta$ is the cone half angle of the jet,  defines the angular region of the jet. 
At $ \mu^2_{\delta}$ we determine the mixing of the three and two jet operators. 
We combine these results with the running of the two jet shape function, which we run down to an energy cut scale $\mu^2_{\beta}$.
This defines the resumed SW dijet decay rate in the context of SCET.
The approach outlined here demonstrates how to establish a jet definition in the context of SCET. This
allows a program of systematically improving the theoretical precision 
of jet phenomenology to be carried out.
\end{abstract}

\maketitle

\section{Introduction}
The phenomenology of QCD jets has been examined over the years with
increasingly sophisticated techniques
\cite{SW,Dokshitzer:1978hw,ellispetronzio,collinssoper,mukhisterman,matrixelementshower}.
Recently, an effective field theory (EFT) of QCD containing collinear
degrees of freedom interacting with soft and ultrasoft gluons, $\rm SCET_{i}$ \cite{SCETlukebauer01},
was formulated.
Using SCET, nonperturbative corrections to dijet event shapes have been examined in a series of
papers \cite{SCETJet1,SCETJet2,SCETJet3,SCETJet4}. 

In this paper, we use SCET to sum large phase space logarithms\footnote{
It has been shown that phase space logs can 
to be resumed for the decay $\btoc$ in \cite{Bauer:1996ma}.} present in the
perturbative expansion of the dijet decay of the $\rm Z$ boson.  Our approach is general and addresses  the problem of establishing a jet definition within SCET.
We have chosen our initial state to be a $\rm Z$ boson to avoid complications from $\rm QCD$ interactions with the initial state.
The logarithms resummed 
have the form $\alpha_s^n \, \log^{n+1}$
in fixed order perturbation theory. At ${\mathcal{O}}(\alpha_s)$ we are resumming the $\alpha_s \, \log \left(2 \, \beta \right)  \, \log \left(\delta \right)$ double log in terms of the jet parameters $\delta$ and $\beta$. We also resum a class of  $\alpha_s^n \, \log^{n}$ logarithms given by $\alpha_s^n \, \log^n \left(\delta \right)$ terms in the fixed order perturbative expansion.
Here $\delta$ and $\beta$ 
are parameters that impose phase space cuts according to the SW jet definition.\footnote{The {\rm SW} jet definition is not the only possible definition of a
jet. In fact, $\rm SW$ jets are somewhat experimentally disfavored compared to the 
$k_{\perp}$ algorithm \cite{Catani:1991hj,tkachov}. The relative theoretical simplicity of the SW jet definition motivates us to demonstrate the origin of 
the relevant logarithms using this definition.}

The SW jet definition defines the dijet decay rate as an integration of the triple differential decay rate 
over a phase space restricted by cuts. 
The cuts on the energy and angular
integrations define the dijet decay rate to be the sum of three partonic
final states. The first state (SW1), is comprised of a quark and anti-quark contained within different jets, where the jets are angularly defined as
cones of half angle $\rm \delta$.
The second state (SW2), is comprised of 
a quark and anti-quark each contained within a different jet, 
plus a gluon (unrestricted in direction) with energy $E_g < \beta \, M_z$.
The final state (SW3),  is defined as 
a quark and an anti-quark within different jets and a gluon
of energy $E_g > \beta \, M_z$ within the jet cones. \footnote{Note that both $\delta$
and $\beta$ are small ($\ll 1$) in this definition;  however one
must also have $\beta$ less than $\delta$ such that $\sin(\delta)
> \beta/(1- \beta)$\cite{stevenson}.}

In this paper, we match 
the amplitude for the decay of the $\rm Z$ onto an operator basis 
defined in $\rm SCET$. This defines contributions 
of two and three jet final states at the scale $M_z^2$. We
run down from the scale $M_z^2$ to the scale defining the dijet 
through the angular cut on phase space. This cut scale is
given by $\mu^2_\delta =  \delta^2 \, M_Z^2$. 
The usoft degrees of freedom are matched onto   
a dijet shape function as in \cite{SCETJet1, SCETJet2,SCETJet3,SCETJet4}. 
This dijet shape function is restricted to emit gluons that take away energy 
$E_g < \beta \, M_z$. This requires the 
introduction of a further phase space scale corresponding to this energy cut. This 
cut scale is given by $\mu^2_{\beta}  =  \delta^4 \, M_Z^2 / {\mathcal{B}}^{1-\delta}$, where ${\mathcal{B}}$ is an ${\mathcal{O}}(1)$ parameter.
We fix ${\mathcal{B}}$ by matching onto the {\rm SW} dijet decay rate after 
we run down to the jet scales. We discuss the physical significance of these 
jet scales in Section \ref{cuts} in more detail.

The resummation of Sudakov \cite{Sudakov} logs for jets has been investigated 
in \cite{mukhisterman,stermanresum, sterman87, catani89, sterman98,CataniResum} for various jet definitions.
The leading logarithm results reported here have been produced in these other contexts. 
We approach the
re-summation with the EFT formalism as this approach offers several advantages. 
For example, the EFT construction includes nonperturbative effects through the
matrix elements of the usoft degrees of freedom. Within the effective theory, one can calculate
to high orders in RGE improved perturbation theory in a clearly defined
and easy manner.  As an example of the utility
of the effective theory approach, we have determined the mixing of the
three and two jet operators. This mixing determines a class 
of NLL contributions in the RGE improved perturbative expansion.  

It is important to note that although the SCET final states discussed here are an 
inclusive sum over the set of hadronic final states, the sum is not a colour singlet. 
Gluons are exchanged between the low energy partons
comprising the jets in the hadronization process. This is a difference in 
calculating dijet production compared to the use 
of the operator product expansion in other inclusive decays, such as semileptonic $B \rightarrow X_c \, \ell \, \nu$ decay \cite{Trott:2004xc}.
Experimental results \cite{opaljets} indicate that this issue does not 
invalidate the SW jet definition.

The outline of this paper is as follows. In the remaining parts of the introduction,
we calculate the $\orderalpha$ phase space logarithms of the SW jet definition.
We then briefly review $\rm SCET$.  In Section \ref{cuts} we formulate the 
phase space cut scales corresponding to the SW jet definition.
In Section \ref{dijet} we discuss the matching of $\Gamma(Z \rightarrow q \, \bar{q})$ onto the
dijet operator. We renormalize the operator basis and run down to the
the scale $\mu^2_\delta$. 
We match the contribution of collinear gluon emission in $Z \rightarrow q \, \bar{q} \, g$ 
onto three jet operators in Section \ref{trijet} and renormalize the three jet operator basis.
Section \ref{running} is devoted to the running of these operators. In Section
\ref{mixing} the mixing of the three and two jet operators is discussed. In Section \ref{results} we introduce the `Phase Space Renormalization Group' and run to the 
phase space cut scales. Finally, we
obtain the leading log resumed dijet decay rate in SCET.
 
\subsection{Phase space Logarithms at $\mathcal{O}(\alpha_s)$}

First, we review how the phase space logarithms present in the {\rm
QCD} calculation of $\Gamma(Z \rightarrow J \, \bar{J})$ are
generated by the SW jet definition. Recall that to $\orderalpha$ the amplitude for $\Gamma(Z \rightarrow
J \,  \bar{J})$ is given by the following expression, grouped in
terms is $q \, \bar{q}$ and $q \, \bar{q} \, g$ final states
\begin{eqnarray}
{{\cal{A}}^2_{Z_{JJ}}} = \mid{{\cal{A}}_0 + {\cal{A}}_V + {\cal{A}}_{W1} + {\cal{A}}_{W2}}\mid^2 + \mid{{\cal{A}}_{B1} + {\cal{A}}_{B2}}\mid^2_R.
\end{eqnarray}
The subscript $\rm R$  refers to the restricted set of  $q \,
\bar{q} \, g$ final states in the jet definition. The amplitudes above are the tree
level amplitude ${\cal{A}}_0$, the vertex correction amplitude ${\cal{A}}_V$ and the 
wavefunction amplitudes ${\cal{A}}_{Wi}$. The ${\cal{A}}_{Bi}$ are the bremstraulung
amplitudes.

Consider first ${\cal{A}}_0$ and the corrections $ {\cal{A}}_V, {\cal{A}}_{W1},{\cal{A}}_{W2} $. 
These contributions to $\Gamma(Z \rightarrow q \, \bar{q})$ can be determined by calculating the imaginary part of
the forward scattering diagrams in Fig. (1).
\begin{figure}[htbp]
\centerline{\scalebox{1.2}{\includegraphics{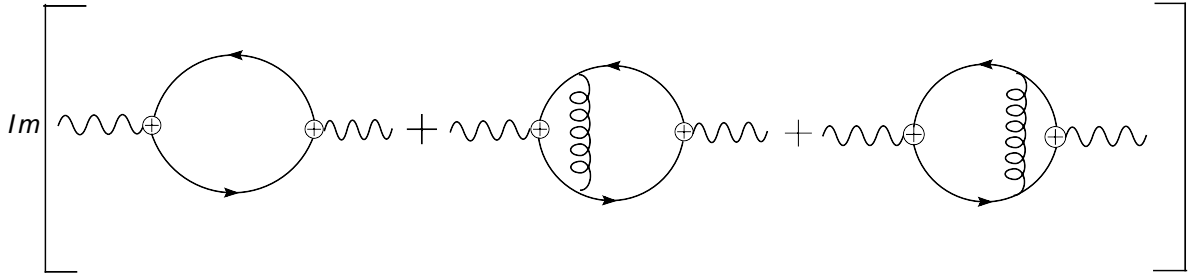}}}
\caption{The imaginary part of the forward scattering amplitude in QCD gives the $q \, \bar{q}$  final state. The operator insertion is the weak neutral current $j_{nc} =  \bar{q} \, \Gamma^{\sigma} \, q \, \epsilon_{\sigma}$ with $\Gamma^{\sigma} = \gamma^{\sigma} \, (g_V + g_A \, \gamma_5)$.  The wavefunction renormalizaton graphs are not drawn as Landau gauge can be used so that the wavefunction contributions vanish. The sum of imaginary part of the diagrams is gauge independent at $\orderalpha$. }
\end{figure}
We calculate in the massless quark limit throughout this paper in the rest frame of the $\rm Z$ , and always  in $d= 4 - 2 \, \epsilon$
dimensions. We average over the initial polarization of the $\rm Z$ in
$d-1$ dimensions, and obtain the
decay rate
\begin{eqnarray} \label{2jetqq}
\Gamma(Z \rightarrow q \bar{q})(\mu) = \frac{N_C}{32 \, \pi^2}(g_V^2 + g_A^2)[M_Z^{1-2 \,\epsilon} (4 \pi)^{2 \,\epsilon} \frac{2- 2 \,\epsilon}{3 - 2 \,\epsilon} \Omega_{3-2 \, \epsilon} ] C^{\overline{\mathrm{MS}}}_{q \, \bar{q}}(\mu).
\end{eqnarray}
We have presented our results in the form suggested in \cite{
SCETJet2} and agree with their result. The coefficient in $\rm
{\overline{MS}}$ is
\begin{eqnarray}\label{tbody}
C^{\overline{\mathrm{MS}}}_{q \, \bar{q}}(\mu) = 1 + \frac{\alpha_s(\mu) \, C_F}{\pi} \left((\frac{1}{\epsilon} + \frac{3}{2})( \ln{[\frac{-2 p_q\cdot p_{\bar{q}}}{\mu^2}] - \frac{1}{\epsilon} }) - 4 + \frac{ \pi^2}{12}
- \frac{1}{2}\ln^2{[\frac{-2 p_q \cdot p_{\bar{q}}}{\mu^2}]} \right) + \orderalphasqr.
\end{eqnarray}
The $q \, \bar{q} \, g$ final state amplitude is $
{\cal{A}}_{Z_{q\bar{q}g}} = \mid{{\cal{A}}_{B1} +
{\cal{A}}_{B2}}\mid^2 $  and is obtained by calculating the
bremstraulung graphs. The double
differential rate is

\begin{figure}[htbp]
\centerline{\scalebox{1.0}{\includegraphics{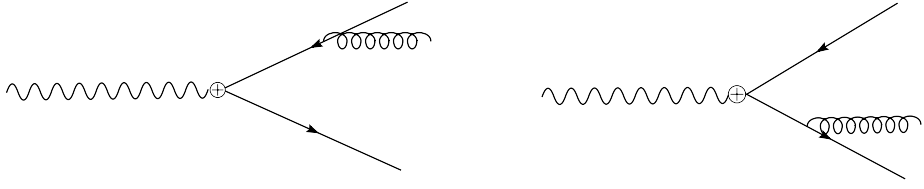}}}
\caption{The three body $q\, \bar{q} \, g$  contributions.}
\end{figure}

\begin{eqnarray} \label{2jetqqg}
\frac{d \Gamma_{q\, \bar{q} \, g}}{d x_q \, d x_{\bar{q}} } =
\frac{ N_C \, C_F \, (g_V^2 + g_A^2)}{256 \, \pi^4} (M_z^{1-2 \, \epsilon}\frac{(4 \, \pi)^{4 \, \epsilon} \, \Omega_{3- 2 \,\epsilon} \, \Omega_{2- 2 \,\epsilon}}{(3 - 2 \,\epsilon)}) x_q^{(d-4)} \, x_{\bar{q}}^{(d-4)} (1 - z^2)^{(\epsilon)} \, F[x_q, x_{\bar{q}}, \mu],
\end{eqnarray}
where the energies of the decay product quarks are $ x_q = 2 \, E_q / M_z$ and  $x_{\bar{q}} = 2 \, E_{\bar{q}} / M_z$. The angle between the quarks has been fixed by a delta function to be
$z = 1 +2 ( 1- x_q - x_{\bar{q}})/ (x_q \, x_{\bar{q}})$.
The function F is kept to order $\epsilon^2$ and is given by
\begin{eqnarray}
F[x_q, x_{\bar{q}}, \mu] = \alpha_s(\mu) (\frac{\mu}{M_Z})^{(1- 2 \, \epsilon)} \large[ (1+ \epsilon)^2 {\frac{x_q + x_{\bar{q}}}{(1- x_q)(1 - x_{\bar{q}})}}+ 2 \,\epsilon (1+ \epsilon)(\frac{2(1- x_q - x_{\bar{q}})+ x_q \, x_{\bar{q}}}{(1- x_q)(1 - x_{\bar{q}})}].
\end{eqnarray}
Integrating this expression over the remaining three body phase space in  $d$ dimensions gives
\begin{eqnarray}
\Gamma(Z \rightarrow q \bar{q} g)(\mu) = \frac{N_C}{32 \, \pi^2}(g_V^2 + g_A^2)[M_Z^{1 - 2 \, \epsilon} (4 \pi)^{2 \, \epsilon} \frac{2- 2 \, \epsilon}{3 - 2 \, \epsilon} \Omega_{3 - 2 \, \epsilon} ] C^{\overline{\mathrm{MS}}}_{q \, \bar{q} g}(\mu),
\end{eqnarray}
with
\begin{eqnarray}
C^{\overline{\mathrm{MS}}}_{q \, \bar{q} g}(\mu) =  - \frac{\alpha_s(\mu) \, C_F}{ \pi} \left((\frac{1}{\epsilon} + \frac{3}{2})( \ln{[\frac{M_Z^2}{\mu^2}] - \frac{1}{\epsilon} }) - \frac{19}{4} + \frac{7 \,  \pi^2}{12}
- \frac{1}{2}\ln^2{[\frac{M_Z^2}{\mu^2}]} \right) + \orderalphasqr.
\end{eqnarray}

The full decay rate to $\orderalpha$ is the result
\begin{eqnarray}
\Gamma_Z(M_z) = \frac{N_C \, M_z (g_V^2 + g_A^2)}{12 \, \pi}\left(1 + \frac{3 \, \alpha_s(M_z) \, C_F }{4 \, \pi} \right) + \orderalphasqr.
\end{eqnarray}

The SW jet definition introduces cuts on the
energy and angular integrations of Eq (\ref{2jetqqg}). The resulting  
phase space integrations are difficult, some details can be found in \cite{SCETJet2}. 
Retaining the terms that diverge as the energy fraction $\beta$ and the cone angle
$\delta$ go to zero,  one obtains
\begin{eqnarray}
C^{SW \, cuts}_{q \, \bar{q} g}(\mu) &=&  - \frac{\alpha_s(\mu) \, C_F}{\pi} \left((\frac{1}{\epsilon} + \frac{3}{2})( \ln{[\frac{M_Z^2}{\mu^2}] - \frac{1}{\epsilon} }) 
- \frac{1}{2}\ln^2{[\frac{M_Z^2}{\mu^2}]} \right) \\
&\,& + \frac{\alpha_s(\mu) \, C_F}{\pi} \left(- 4 \ln (2 \, \beta) \ln (\delta) - 3 \ln (\delta) + \frac{13}{2} - \frac{11 \, \pi^2}{12} \right).
\end{eqnarray}

Combining this result with Eq.(\ref{tbody}), one obtains the SW
result;  we find 
\begin{eqnarray}\label{sterW}
\Gamma(Z \rightarrow J \bar{J})(M_z) = \frac{N_C \, M_z (g_V^2 + g_A^2)}{12 \, \pi}\left(1 + \frac{\alpha_s \, C_F }{\pi} (\frac{5}{2} - \frac{\pi^2}{3}- 3 \ln \delta - 4 \ln (2 \, \beta) \ln (\delta) ) \right).
\end{eqnarray}
Note that the result is free of explicit $\rm IR$ singularities. This is expected as the SW jet definition is constructed to satisfy the $\rm KLN$ theorem \cite{LeeNauenberg,kinoshita}.
However, the result does depend on logs of the cuts used to partition phase space.

We will now examine this decay in $\rm SCET$. 
The different final state phase space configurations
will be separated out into different operators using the invariant $p_g \cdot p_{i}$. 
By matching and running, the renormalization group allows us 
to sum the Sudakov logarithms and define a dijet amplitude.

\subsection{Jets in SCET}
$\rm SCET$ contains collinear and ultrasoft degrees of freedom 
that are relevant for $\rm QCD$ jet studies.\footnote{ See \cite{SCETlukebauer01} for more detailed reviews.}
Consider a final state gluon produced by the decay products of the Z.  The gluon is collinear in direction $n_i$ when it has
scaling $p_c^{(n_i)} = M_Z ( \lambda^2, 1, \lambda)$, where $\lambda$ is a small parameter in the effective theory.
We have used the lightcone coordinates $L^{\mu} = (L^+, L^-,L_{\perp})$ such that $L^+ = n_i \cdot L$,$L^- = {\bar{n}}_i \cdot L$ and ${L_i}_{\perp}^{\mu} = L^{\mu} - L^+ \, {\bar{n}}_i^{\mu} /2 -  L^- \, {n_i}^{\mu} /2$.
The lightcone vectors used here are defined as  $n_i = (1,\vec{n}_i)$ and $\bar{n}_i = (1,-\vec{n}_i)$ with $\vec{n}_i$ a unit vector.

Alternatively, the gluon could have ultrasoft scaling $ p_{us} = M_Z ( \lambda^2, \lambda^2, \lambda^2)$. When 
$\lambda \sim \sqrt{\LQCD/ M_Z}$,  one cannot perturbatively expand the decay rate in terms of usoft gluons. The nonperturbative effects of these gluons are given by the matrix elements of ultrasoft degrees of freedom, which give
nonperturbative corrections to the decay rate \cite{SCETJet1,
SCETJet2,SCETJet3,SCETJet4}.
For this reason, we take our final state gluon radiation 
to be collinear in our perturbative expansion.
Expanding the decay current  $j^{\mu}$ in terms of collinear and ultrasoft  degrees of freedom, the
SCET jet fields are given by
\begin{eqnarray}
 {\bar {\chi}}_{n_q} &=& [\bar{\xi}_{n_q} \, W_{n_q} ], \nn \\
 {\chi}_{n_{\bar{q}}} &=&  [ W^{\dagger}_{n_{\bar{q}}} \xi_{n_{\bar{q}}}].
\end{eqnarray}
The ${\xi}$ field is defined with the large momenta $\tilde{p}$ in the direction $n_{i}$ removed
\begin{eqnarray}
\xi_{n_i}(x) = \sum_{\tilde{p}} e^{- i \, {\tilde{p}} \cdot x} \,{P}_{n_{i}}\psi(x),
\end{eqnarray}
using the projector
\begin{eqnarray}
 {P}_{n_{i}} = \frac{n_i \! \! \! \! \! / \, \, {\bar{n}}_i \! \! \! \! \! /}{4}.
\end{eqnarray}
Also present to preserve collinear gauge invariance are Wilson lines,
\begin{eqnarray}
W_{n_i}(q) = \left[ \sum_{perms} \exp \left(- \frac{1}{ {\overline{\cal{P}}}} \, \bar{n}_i \cdot A_{n_i,q}\ \right) \right],
\end{eqnarray}
\begin{eqnarray}
W^{\dagger}_{n_i}(q) = \left[ \sum_{perms} \exp \left(-  \bar{n}_i \cdot A^{\star}_{n,q}\ \right) \frac{1}{ {\overline{\cal{P}}^{\dagger}}}\right].
\end{eqnarray}

\section{The Scales of the SW Cuts}\label{cuts}

Three small parameters exist in the construction of $\rm SW$ jets $\LQCD/M_Z, \delta$ and $\beta$.
We take $\delta \gg \sqrt{\LQCD/M_Z}$ for the cuts to be above the hadronization scale. To sum large phase space logs with the renormalization group, we must associate the phase space cuts with scales  $\mu^2_{cut}$.  

For the angular region defining the jet ($\delta$), we utilize the approach of \cite{Dokshitzer:1978hw,ellispetronzio,sterman95} and take
a collinear momenta comprising the jet to have
collinear scaling  $P_J^{(n_i)} \sim M_Z ( \delta^2, 1, \delta)$. This gives the cut scale of 
\begin{eqnarray}
\mu^2_{\delta} = \delta^2 \,  M_Z^2. 
\end{eqnarray}
We set the invariant mass of the initial off shell quark and anti-quark to this scale, $P_{1,2}^2 =  \delta^2 \,  M_Z^2$.
Physically this scaling has a simple interpretation. The jet cone is defined by the half angle $\delta$.  
For all momenta making up the jet,when $z$ is taken as the axis of the jet cone,
\begin{eqnarray}
\tan \left( \delta \right) \sim \frac{|p_{\perp}|}{|p_z|}.
\end{eqnarray}

The usoft gluons radiate into the final state and deposit  a small fraction of center of mass energy, $\beta \, M_Z$, isotropically in the detector. For these gluons, we utilize the fact that this  radiation cannot disturb the back to back configuration of the jet cones. 
The sum of usoft gluon emissions is restricted to carry a total energy ($P_u^0$) out of the jet cones  that
satisfies $\sin(\delta)> P_u^0/(1- P_u^0)$ \cite{stevenson}. This
requires the momenta emitted outside the jet cones to have scaling $P_u \sim M_Z ( \delta^2,  \delta^2, \delta^2)$. We take
the energy cut scale of the usoft degrees of freedom to be
\begin{eqnarray}\label{beta}
\mu^2_{\beta} =  {\mathcal{B}}^{\delta -1} \,  \frac{\mu^4_{\delta}}{M_Z^2},
\end{eqnarray}
where ${\mathcal{B}} $ is an ${\mathcal{O}}(1)$ parameter. 
The correct resummation of leading logarithms in the SW jet definition requires that $\mu^2_{\beta}/\mu^2_{\delta} \propto \delta^2$. Physically, this corresponds to the usoft radiation not being able to 
cause the quark and anti-quark to lie outside the jet cones. 
Note that the results of the SW phase space integrals in Eq.(\ref{sterW}) are expanded in $\delta$ and $\beta$. All dependence on the jet parameters $\delta$ and $\beta$ that does not diverge as $\beta, \delta \rightarrow 0$ is neglected in Eq.(\ref{sterW}). 
In a similar manner, the dependence on ${\mathcal{B}}, \delta$ in our RGE is expanded.  
Dependence on these parameters that does not lead to large logs as 
$\beta, \delta \rightarrow 0$ is neglected. In this manner, any
$\mu^2_{\beta}$ dependence on ${\mathcal{B}}$ that gives the same retained
contribution to the RGE will lead to the same double Sudakov logarithm. 
The quark and anti-quark ${\mathcal{B}}$ dependence is sub leading in the parameter $\delta$ and neglected except for the component that scales as $\delta^2$.
How the particular components of the usoft gluons depend on ${\mathcal{B}}$
is fixed when we match onto the SW jet definition. 
We will demonstrate how these phase space scales will establish the SW jet definition in SCET in Section \ref{PSRG}.

\section{$Z \rightarrow q \, \bar{q} $ in SCET}\label{dijet}
At the scale $\mu^2 = M_z^2$ we match $\Gamma(Z \rightarrow q \, \bar{q})$ in QCD onto the dijet operator \cite{aneeshDIS}
\begin{eqnarray}
O_2^{\sigma} = {\bar{\chi}}_{n_q} \, \Gamma^{\sigma} \, {\chi}_{n_{\bar{q}}}.
\end{eqnarray}
The EFT loops are scaleless and vanish in dimensional reqularization at $\orderalpha$. The Wilson coefficient $C_{2}$
and the remormalization factor $Z_{2}$ depend on the large label momenta components $2 \, p_q\cdot p_{\bar{q}} \sim M_Z^2$ of the
quark and anti-quark \cite{aneeshDIS}. The  Wilson coefficient is given by the one loop matching condition
\begin{eqnarray}\label{match}
\langle q \, {\bar{q}} |{\cal{A}} | Z_\sigma \rangle(\mu)  = \frac{\langle O_2^{\sigma}  \rangle (\mu)  \, C_{2}(\mu)}{ Z_{2}(\mu) },
\end{eqnarray}
determining
\begin{eqnarray}
C_{2}(\mu) &=& 1 + \frac{\alpha_s(\mu^2) \, C_F}{2 \, \pi} ( \frac{3}{2} \ln{[\frac{-2 p_q\cdot p_{\bar{q}}}{\mu^2}]}  - 4 + \frac{\pi^2}{12}
- \frac{1}{2}\ln^2{[\frac{-2 p_q \cdot p_{\bar{q}}}{\mu^2}]} ) + \orderalphasqr,  \nn \\
Z_{2}(\mu)&=& 1 + \frac{\alpha_s(\mu^2) \, C_F}{2 \, \pi} ( \frac{1}{\epsilon^2} + \frac{3}{2 \,\epsilon} - \frac{1}{\epsilon}  \ln{[\frac{-2 p_q\cdot p_{\bar{q}}}{\mu^2}]} ) + \orderalphasqr.
\end{eqnarray}

The contribution from this operator at the scale $\mu^2 = M_Z^2$ is given by Eq.(\ref{match})
with 
\begin{eqnarray}
C_2(M_Z) = 1 + \frac{\alpha_s(M_Z) \, C_F}{2 \, \pi}\left( 3 \, i -4 + \frac{7 \pi^2}{12}\right) + \orderalphasqr.
\end{eqnarray}
 Note that contribution to the decay rate from the imaginary part of the Wilson coefficient vanishes at $\mathcal{O}(\alpha_s)$. The imaginary term can contribute at $\orderalphasqr$.

To determine the counter terms in the effective theory, we use offshellness to regulate the IR.  Both soft and collinear loops contribute to the UV divergences that determine the counter term. If one adopts two stage running\footnote{We find it convenient to adopt one stage running in Section \ref{PSRG}. However we utilize two stage running
initially to demonstrate the benefit of one stage running in Section \ref{PSRG}.}, the subtraction points of the usoft and collinear
loops are taken to be the same, ie $\mu^2_s = \mu^2_c$. With this choice, the anomalous dimension of $\rm O_2$
is given by \cite{aneeshDIS}
\begin{eqnarray}\label{anomdimalpha}
\gamma_2 &=& - 2 \alpha_s \frac{d Z_{2}^\epsilon(\alpha_s)}{d \alpha_s}, \nn \\
&=& -\frac{\alpha_s(\mu)}{\pi} \, C_F \, (\log \left(\frac{\mu^2}{M_Z^2} \right)+ \frac{3}{2} )  + \orderalphasqr.
\end{eqnarray}
Note that we have used the $\rm \overline{MS}$ result \cite{Floratos} for the epsilon poles 
of the counter term $Z_2^\epsilon$.

The Wilson coefficient of the operator at this lower scale is obtained
by solving the RGE, with $\beta_0 = 11 - 2 n_f /3$. For two stage running 
we find
\begin{eqnarray}
C_2^{two}(\mu) =  \left(\frac{\alpha_s(\mu^2)}{\alpha_s(M_Z^2)} \right)^{\frac{C_F}{\beta_0}\left(3 - \frac{8 \, \pi}{\beta_0 \, \alpha_s(M_Z^2)}\right)} \, \left(\frac{\mu^2}{M_Z^2}\right)^{\left(\frac{- 2 \, C_F}{\beta_0} \right)} C_{2}(M_Z).
\end{eqnarray} 

\section{$Z \rightarrow q \, \bar{q} \, g$ in SCET}\label{trijet}

Beyond tree level, we match the contribution of  $\Gamma(Z \rightarrow q \bar{q} g)$ onto
the two and three jet operator basis. The QCD amplitude is matched onto a two jet operator 
when the collinear gluon emission can be described by ${\mathcal{L}}_{SCET}$ and 
$P^{2}_{qg} < \delta^2 \, M_Z^2$ or $P^{2}_{\bar{q}g} < \delta^2 \, M_Z^2$,
where $ P^{\alpha}_{ig} = P^{\alpha}_i + P^{\alpha}_g $. 
Conversely, when $P^{2}_{ig} \ge \delta^2 \, M_Z^2$ for both $i = q,\bar{q}$, the collinear gluon emission is matched onto a three jet
operator. At the scale $\mu^2 = M_z^2$, we have the following amplitude depicted in Fig. (2)
\begin{eqnarray}
{{\cal{A}}_{B1} + {\cal{A}}_{B2}} = \bar{u}(p_q)^{s_2} \left[(- i g_s \gamma_{\nu} \, T^a) \left(i  \frac{P_{qg}  \! \! \! \! \! \! \! /}{P_{qg}^2} \right) \Gamma^{\mu} -  \Gamma^{\mu}
\left(i  \frac{P_{{\bar{q}}g}  \! \! \! \! \! \! \! /}{P_{{\bar{q}}g}^2} \right)(- i g_s \gamma_{\nu} \, T^a) \right] v(p_{\bar{q}})^{s_1} \epsilon'_{\mu} \, \epsilon^{\nu}.
\end{eqnarray}
For the three jet operators,  we take $2 \,  p_g \cdot p_{i}, \sim  M_Z^2$ and the intermediate propagator is not resolved between the scales $M_Z^2$ and $\delta^2 \, M_Z^2$.
Our matching to order
$\lambda^0 M_Z^2$ is given by 
\begin{eqnarray}
{{\cal{A}}_{B1} + {\cal{A}}_{B2}} =  \sum_{i = 3,4,5}\, \frac{C_{i}(\mu)}{Z_{i}(\mu)} \langle q \, {\bar{q}} \,  A_c^{\nu} |  O_i^{\sigma} | \epsilon_\sigma \rangle (\mu),
\end{eqnarray}
where we have suppressed the summations over the light cone directions $n_i$. For different phase space configurations,  $2 \,  p_g \cdot p_{i}$ takes on different values 
between $M_Z^2$ and $\delta^2 \, M_Z^2$. In running between $M_Z^2$ and $\delta^2 \, M_Z^2$ the logs generated by the separation of scales
depend on the phase space configuration of the event. Our formalism takes this into account by $n_i$ dependence in the anomalous dimensions of the three jet operators.

To decompose the amplitude into operators with field content defined in $\rm SCET$, 
we first rearrange the three parton amplitude into the form
\begin{eqnarray}
{{\cal{A}}_{B1} + {\cal{A}}_{B2}} &=& \bar{u}^{s_2}(p_q) \left[ \frac{P_{\bar{q}}^{\alpha} \, P_q^{\beta} \,
({P_g}_{\alpha} \, \epsilon_\beta^a -  \epsilon_\alpha^a  \, {P_g}_{\beta}) }{(P_g \cdot P_q)(P_g \cdot P_{\bar{q}})} \right] \, \Gamma^{\sigma} ( g_s \, T_a) \, v^{s_1}(p_{\bar{q}}) \, \epsilon'_{\sigma} \nn \\
&+&  \bar{u}^{s_2}(p_q) \left[\frac{P_g \cdot (P_{\bar{q}} - P_q) ({P_g}_{\sigma} \, \epsilon_\beta^a -  \epsilon_\sigma^a  \, {P_g}_{\beta}) }{2\, (P_g \cdot P_q)(P_g \cdot P_{\bar{q}})}
\right] \, \Gamma_{\beta} ( g_s \, T_a) \, v^{s_1}(p_{\bar{q}}) \, \epsilon'_{\sigma} \nn \\
&+&  \bar{u}^{s_2}(p_q) \left[\frac{P_g \cdot (P_{\bar{q}} - P_q)\, i \, \epsilon^{\alpha \, \beta \, \sigma \, \eta } \epsilon_{\alpha}^a \, \epsilon'_\sigma \, {P_g}_{\beta} \, \gamma_5}{2\, (P_g \cdot P_q)(P_g \cdot P_{\bar{q}})}\right] \, \Gamma_{\eta} ( g_s \, T_a) \, v^{s_1}(p_{\bar{q}}).
\end{eqnarray}
We now expand such that the partons are collinear particles in the directions $n_1, \, n_2, n_3$ for $p_q$, $p_{\bar{q}}$ and $p_g$ respectively. 
The operator basis in $\rm SCET$ is  dictated by the constraint of collinear gauge invariance in the three directions $n_1,n_2, n_3$. We find the following operators,
\begin{eqnarray} \label{ops}
O_3^{\sigma} &=&  {\bar{\chi}}_{n_q}  \, {\mathcal{C}}_3^{\alpha \, \beta}(\bar{n}_3 \cdot p_3,n_i) \, [W^{\dagger}_{n_3} \,g_s \,  G^{n_3}_{\alpha, \beta} \, W_ {n_3}   ] \, \Gamma^{\sigma} {\chi}_{n_{\bar{q}}}, \nn \\
O_4^{\sigma} &=& {\bar{\chi}}_{n_q}\, {\mathcal{C}}_4( \bar{n}_i \cdot p_i ,n_i) \, [W^{\dagger}_ {n_3}  \,  g_s \, G_{n_3}^{\alpha, \sigma} \, W_ {n_3}   ] \, \Gamma_{\alpha}\, {\chi}_{n_{\bar{q}}}, \nn \\
O_5^{\sigma} &=& {\bar{\chi}}_{n_q}\, {\mathcal{C}}_5( \bar{n}_i \cdot p_i,n_i) \, [W^{\dagger}_ {n_3} \, g_s \, {\tilde{G}_{n_3}}^{\sigma, \eta} \, \gamma_5 \, W_ {n_3}   ] \, \Gamma_{\eta} \, {\chi}_{n_{\bar{q}}}.
\end{eqnarray}
The Wilson coefficients are of the form
\begin{eqnarray}
{\mathcal{C}}_3^{\alpha \, \beta}(\mu, \bar{n}_3 \cdot p_3,n_i) &=&\frac{4 \, n_1^\beta \, n_2^\alpha}{(n_1 \cdot n_3) \, (n_2 \cdot n_3) ({\bar{n}}_3 \cdot p_3)^2 } \, C_3 \left(\mu \right),  \\
{\mathcal{C}}_{4,5} (\mu, \bar{n}_i \cdot p_i ,n_i) &= & \left( \frac{2}{(n_1 \cdot n_3)  \, (\bar{n}_1 \cdot p_1) \, (\bar{n}_3 \cdot p_3)} 
 - \frac{2}{(n_2 \cdot n_3) \, (\bar{n}_2 \cdot p_2) (\bar{n}_3 \cdot p_3)} \right)
\,  C_{4,5} \left(\mu \right) \nn
\end{eqnarray}
where the non label part of the Wilson coefficient is 
\begin{eqnarray}
C_i \left( \mu \right) = \left[C_i^0+ \alpha_s(\mu) C_i^1(\mu) + \orderalphasqr \right],
\end{eqnarray}
with $C_i^0 = 1$.
We have used the ${\mathcal{O}}(\lambda^0)$ collinear gluon field strength  $g_s \,G^{\alpha, \beta}_{n_3} \equiv  \, [ i {\cal{D}}^\alpha_{n_3} +  g_s \, A^\alpha_{n_3,p_g},  i {\cal{D}}^\beta_{n_3} +  g_s \, A^\beta_{n_3,p_g}] $ and
$  {\tilde{G}}_{n_3}^{\sigma, \eta} =  i \, \epsilon^{\alpha \, \beta \, \sigma \, \eta } G^{n_3}_{\alpha, \beta} $
is the corresponding dual field. The amplitude expressed in terms of these operators requires a summation over lightcone directions. This summation is such that when
$n_i$ and $n_j$ label distinct collinear directions, they lie in different jet cones. This constraint along with momentum conservation is
\begin{eqnarray}
\hat{C}_3(\mu^2) =\sum_{n_i} C_3(\mu^2) \, \theta \left[n_i \cdot \bar{n}_j  - (1+ \cos{2 \delta}) \right] \, \delta( n_1 \cdot n_2 + n_1 \cdot n_3 + n_2 \cdot n_3 - 2). 
\end{eqnarray} 
We suppress these summations and constraints until Section \ref{mixing}.   

The three jet operators contain the same field content. In $\rm SCET$
these operators are expected to have the same renormalization
\cite{BauerSchwartz, Conversations}. We present the operators in the form of Eq.(\ref{ops}) as they have a clear field 
content and Dirac structure. We also find this form convenient for the calculations in Section \ref{mixing}.

A collinear gluon with large angular separation from both the quark and anti-quark can also 
be produced when the quark and anti-quark are collinear. In the original calculation of SW, these states did 
not contribute at $\orderalpha$. In the effective field theory, the corresponding operators vanish at leading order in 
$\lambda$. This is another example of the the utility
of SCET.

\subsection{Renormalization of the 3-Jet Operators}\label{renormalize3}

The three jet operators are renormalized by determining the $\rm
UV$ divergences in SCET using the Feynman rules stated in the Appendix.
The $ \rm UV$ divergences are regulated using dimensional regularization
and $\rm \overline{MS}$. We utilize offshellness to separate out
the $\rm IR$ divergences.\footnote{We have used Feynman gauge.}
For some related details on one gluon external state
renormalization calculations, see \cite{TrottWilliamson}. We also note that we
use the background field method \cite{abbott, SCETlukebauer01}.

As we examine perpendicular polarized collinear gluons in
the $n_3$ direction, this causes the contributions  from a number of
diagrams to vanish in the effective theory.
The remaining diagrams are shown in Figure 3.

\begin{figure}[htbp]\label{renorm}
\centerline{\scalebox{1.0}{\includegraphics{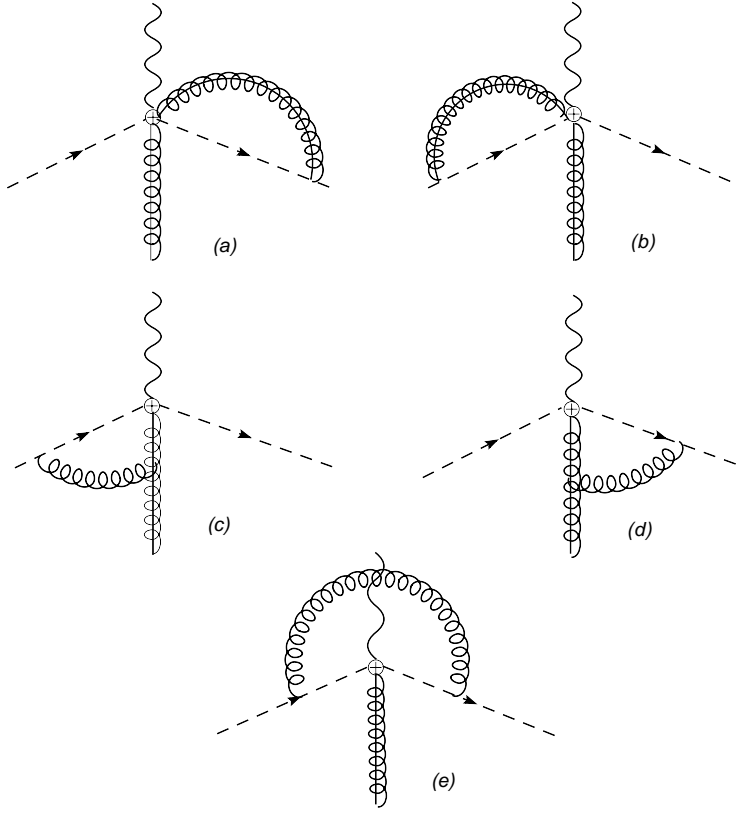}}}
\caption{One gluon external state diagrams for the operators $O_i$ in the effective theory. Collinear gluons are drawn as springs with lines, usoft gluons
as springs.}
\end{figure}

Calculating in the rest frame of the $\rm Z$,  the perturbative expansion of the operator can be  determined. We obtain the following for diagrams (a) and (b)
\begin{eqnarray}\label{diagrama}
{\langle {\xi}_{n_1} \, {\bar{\xi}}_{n_2} \, A^{\perp, \nu}_{n_3}|  \mathcal{O}^{\sigma}_{3} |0 \rangle}_{a}
&=& - \frac{g_s^2 \, C_F\, C^{\nu \, \sigma}}{8 \, \pi^2} \left[-\frac{1}{\epsilon_{\rm uv}^2} -\frac{1}{\epsilon_{\rm uv}}
+ \frac{1}{\epsilon_{\rm uv}} \, \log \left[ \frac{- P_{q}^2 }{ \mu^2} \right] \right] \nn \\
&\,&- \frac{g_s^2 \, C_F \, C^{\nu \, \sigma}}{8 \, \pi^2} \left[  - \frac{1}{2}  \log^2 \left[ \frac{- P_{q}^2 }{ \mu^2} \right]
+  \log \left[ \frac{-P_{q}^2 }{ \mu^2} \right] -2 + \frac{\pi^2}{12} \right],
\end{eqnarray}
\begin{eqnarray}
{\langle {\xi}_{n_1} \, {\bar{\xi}}_{n_2} \, A^{\perp, \nu}_{n_3}|  \mathcal{O}^{\sigma}_{3} | 0 \rangle}_{b}
&=& - \frac{g_s^2 \, C_F\, C^{\nu \, \sigma}}{8 \, \pi^2} \left[-\frac{1}{\epsilon_{\rm uv}^2} -\frac{1}{\epsilon_{\rm uv}}
+ \frac{1}{\epsilon_{\rm uv}} \, \log \left[ \frac{- P_{\bar{q}}^2 }{ \mu^2} \right] \right] \nn \\
&\,&- \frac{g_s^2 \, C_F \, C^{\nu \, \sigma}}{8 \, \pi^2} \left[  - \frac{1}{2}  \log^2 \left[ \frac{- P_{\bar{q}}^2 }{ \mu^2} \right]
 + \log \left[ \frac{- P_{\bar{q}}^2 }{ \mu^2} \right] -2 + \frac{\pi^2}{12} \right].
\end{eqnarray}
We have defined $C^{\nu \, \sigma} = \langle \bar{\xi}_{n_1} \,  \xi_{n_2} \,  A^{\perp, \nu}_{n_3} |  O_3^{\sigma}| 0 \rangle$ in the above.
Defining the following function
\begin{eqnarray}
K \left[(P_i,n_i), (P_j, n_j)\right] & \equiv &
 \log \left[\frac{- 2 \, P_i \cdot P_j}{\mu^2} \right]  -  \log \left[\frac{-P_j^2}{\mu^2} \right] -  \log \left[\frac{-P_i^2}{\mu^2} \right] \nn \\
&\,& - \log \left[ 1 + \frac{n_i \cdot {\bar{n}}_j}{2} \left(\frac{\bar{n}_j \cdot P_j}{\bar{n}_i \cdot P_{i}} \right) \left(\frac{P_i^2}{P_j^2}\right) \right],
\end{eqnarray}
the loop results for $(c,d,e)$ can be expressed as functions of a master loop integral that gives the following form
\begin{eqnarray}
I_1 \left[(P_i,n_i), (P_j,n_j)\right] &=& \int \left(\frac{d \, k}{2 \, \pi} \right)^{d}  \frac{1}{k^2 + i \epsilon} \, \frac{1}{P_i^2/ (n_i \cdot P_i) - n_i \cdot k \ + i \epsilon} \,  \frac{1}{P_j^2/ (n_j \cdot P_j) - n_j \cdot k \ + i \epsilon}
\nn \\
&=& \frac{1}{8 \, \pi^2 \, n_i \cdot n_j} \left[\frac{1}{\epsilon_{\rm uv}^2}
+ \frac{1}{\epsilon_{\rm uv}} K \left[(P_i,n_i), (P_j,n_j)\right]  +  \frac{1}{2} K \left[(P_i,n_i), (P_j,n_j)\right]^2 + \frac{\pi^2}{4}\right]. \nn
\end{eqnarray}

Diagrams (c) and (d) give the following results:
\begin{eqnarray}
{\langle {\xi}_{n_1} \, {\bar{\xi}}_{n_2} \, A^{\perp, \nu}_{n_3}|  \mathcal{O}^{\sigma}_{3} | 0 \rangle}_{c}
&=& -  \frac{g_s^2 \, C_A\, C^{\nu \, \sigma} \, n_2 \cdot n_3}{2} I_1 \left[(P_g,n_3), (P_{\bar{q}},n_2)\right], \nn \\
{\langle {\xi}_{n_1} \, {\bar{\xi}}_{n_2} \, A^{\perp, \nu}_{n_3}|  \mathcal{O}^{\sigma}_{3} | 0 \rangle}_{d}
&=& -  \frac{g_s^2 \, C_A\, C^{\nu \, \sigma} \, n_1 \cdot n_3}{2} I_1 \left[(P_g,n_3), (P_q,n_1)\right].
\end{eqnarray}
In obtaining these results, one must properly account for the zero bin to avoid introducing $\rm IR$ poles into
the anomalous dimension. For this purpose, one can utilize the method outlined in \cite{zerobin}.
Diagram (e) can be calculated in a manner similar to \cite{aneeshDIS}
where the anomalous dimension of $O_2^\sigma$ was determined. We obtain
\begin{eqnarray}
{\langle {\xi}_{n_1} \, {\bar{\xi}}_{n_2} \, A^{\perp, \nu}_{n_3}| \mathcal{O}^{\sigma}_{3} | 0 \rangle}_{e}
&=& -  g_s^2 \, (C_F - C_A/2) \, C^{\nu \, \sigma} \, n_1 \cdot n_2  \,  I_1 \left[(P_{\bar{q}},n_2), (P_q,n_1)\right].
\end{eqnarray}
The remaining diagram
(f) can be determined by utilizing the full background field three gluon vertex \cite{abbott} and the two
gluon feynman rule for the operator. For this diagram one finds 
\begin{eqnarray}
{\langle {\xi}_{n_1} \, {\bar{\xi}}_{n_2} \, A^{\perp, \nu}_{n_3}|  \mathcal{O}^{\sigma}_{3} | 0 \rangle}_{f}
&=&  \frac{g_s^2 \, C_A\, C^{\nu \, \sigma}}{16 \, \pi^2} \left[ \frac{2}{\epsilon_{\rm uv}^2}
- \frac{2}{\epsilon_{\rm uv}} \log \left[\frac{-P_g^2}{\mu^2} \right]  +  \log^2 \left[\frac{-P_g^2}{\mu^2} \right]  - \frac{\pi^2}{6} \right] \nn \\
&-&  \frac{g_s^2 \, C_A\, C^{\nu \, \sigma}}{16 \, \pi^2} \left[-2 - \frac{1}{\epsilon_{\rm uv}} + \log \left[\frac{-P_g^2}{\mu^2}\right] \right].
\end{eqnarray}

The remaining graphs are wave function graphs. For a soft gluon on the external quark field, the wavefunction graph
vanishes as $n_i^2 = 0$. The collinear wavefunction renormalization graphs \cite{SCETlukebauer01} must also be added to this result.
For each effective field $\xi$, with momenta $P_i$, we have the wavefunction renormalization factor
\begin{eqnarray}
Z_{\xi}(P_i,\mu) = 1- \frac{\alpha_s(\mu) \, C_F}{4 \, \pi} \left[ \frac{1}{ \epsilon_{\rm uv}} - \log \left(\frac{-P_i^2}{\mu^2} \right) + 1 \right].
\end{eqnarray}

Combining these results and adding the corresponding counter term
leads to the renormalized perturbative expansion to $\orderalpha$. 
The renormalized operator is given by $O_3^{(r)} = \sqrt{Z_q} \, \sqrt{ Z_{\bar{q}}} \, O_3^{(0)}/Z_3$. As we utilize the 
background field method, the combination $g \, A_g$ is not renormalized \cite{abbott}. 
The renormalization factor of the operator is given by
\begin{eqnarray}\label{threejrenorm}
Z_{3}(\mu)
&=&
1 + \frac{\alpha_s(\mu) \, C_F}{4 \, \pi } \left[\frac{2}{\epsilon^2} + \frac{3}{\epsilon}  - \frac{2}{\epsilon}\left( \log\left[-\frac{P_q^2}{\mu^2}\right] + \log\left[-\frac{P_{\bar{q}}^2}{\mu^2}\right]
+ K \left[(P_q,n_1), (P_{\bar{q}},n_2)\right]  \right) \right], \nn \\
&\,&  + \frac{\alpha_s(\mu) \, C_A}{4\, \pi } \left[ \frac{1}{\epsilon}   K \left[(P_q,n_1), (P_{\bar{q}},n_2)\right] - \frac{1}{\epsilon} K \left[(P_{\bar{q}},n_2), (P_g,n_3)\right] 
- \frac{1}{\epsilon} K \left[(P_{q},n_1), (P_g,n_3)\right]  \right], \nn \\
&\,&  + \frac{\alpha_s(\mu) \, C_A}{4 \, \pi } \left[\frac{1}{\epsilon^2} + \frac{1}{\epsilon}  - \frac{2}{\epsilon} \log\left[-\frac{P_g^2}{\mu^2}\right] \right].
\end{eqnarray}
We scale the results
by the one gluon matrix element  $\langle \tilde{O}_3^{(r)} \rangle = \langle {\xi}_{n_1} \, {\bar{\xi}}_{n_2} \, A^{\perp, \nu}_{n_3}|(O^{\sigma}_3)^{(r)}| 0 \rangle / C^{\nu \, \sigma}$.
The perturbative expansion of the operator in the effective theory is
\begin{eqnarray}
\langle \tilde{O}_3^{(r)} \rangle
&=&
1 +  \, \frac{\alpha_s(\mu) \, C_F}{4 \, \pi } \left[ 7 - \frac{5 \, \pi^2}{6} -\frac{3}{2}  \log \left[ - \frac{P^2_{\bar{q}}}{\mu^2} \right] -\frac{3}{2}  \log \left[- \frac{P^2_{q}}{\mu^2}\right]
+ \log^2 \left[ - \frac{P^2_{\bar{q}}}{\mu^2}\right] + \log^2 \left[ - \frac{P^2_{q}}{\mu^2}\right] \right],  \nn  \\
&-& \frac{\alpha_s(\mu) \, C_F}{4 \, \pi } K \left[(P_q,n_1), (P_{\bar{q}},n_2)\right] ^2 + \frac{\alpha_s(\mu) \, C_A}{4 \, \pi } \left(2 - \frac{5 \, \pi^2}{12}  - \log \left[ -\frac{P^2_g}{\mu^2}\right]
+  \log^2 \left[- \frac{P^2_g}{\mu^2}\right]  \right),  \nn \\
&+& \frac{\alpha_s(\mu) \, C_A}{8 \, \pi }\left(-K \left[(P_q,n_1), (P_g,n_3)\right] ^2 - K \left[(P_{\bar{q}},n_2), (P_g,n_3)\right] ^2  + K \left[(P_{\bar{q}},n_2), (P_q,n_1)\right] ^2 \right).
\end{eqnarray}

The anomalous dimension can be determined from the renormalization factor. The scaling for the inner product  $P_i \cdot P_j $ in this operator basis was defined 
to be $2 \, P_i \cdot P_j \sim M_Z^2$. Using this scaling, and taking the quark and anti-quark lightcone vectors along the jet direction, we have
$ \left(\bar{n}_j \cdot P_j \right) / \left(\bar{n}_i \cdot P_{i} \right) \left(P_i^2/P_j^2\right) =  \left(n_i \cdot P_i \right) / \left(n_j \cdot P_{j} \right) $. Using these results, we determine
\begin{eqnarray}
K \left[(P_i,n_i), (P_j,n_j)\right]  &=&  - \log \left[ \frac{\mu_s^2}{M_Z^2}\right] -  \log \left[ - \frac{P^2_{\bar{q}}}{\mu^2} \right] 
-  \log \left[ - \frac{P^2_{q}}{\mu^2} \right]   - L[(P_i,n_i), (P_j,n_j)], 
\end{eqnarray}
where 
\begin{eqnarray}
L[(P_i,n_i), (P_j,n_j)] =  \log\left[\frac{2 \,  n_j \cdot P_j  + n_i \cdot {\bar{n}}_j \, \left(n_i \cdot P_i \right)}{2 \, \left(n_i \cdot {n}_j \right) \, n_j \cdot P_{j}}  \right].
\end{eqnarray}
The anomalous dimension is determined for two stage running to be
\begin{eqnarray}\label{anomdimalphao3}
\gamma_3(\mu, n_i) &=&  - \frac{\alpha_s(\mu) \, C_F}{4 \, \pi} \left[ 4 \log \left[ \frac{\mu^2}{M_Z^2} \right]  + 6 + 4 
L[(P_1,n_1), (P_2,n_2)]  \right] - \frac{\alpha_s(\mu) \, C_A}{4 \, \pi} \left[ 2 \log \left[ \frac{\mu^2}{M_Z^2} \right]  + 2 \right] \nn \\
&\,&  - \frac{\alpha_s(\mu) \, C_A}{4 \, \pi} \left( L[(P_1,n_1), (P_3,n_3)] + L[(P_2,n_2), (P_3,n_3)] - 
L[(P_1,n_1), (P_2,n_2)]  \right). 
\end{eqnarray}

Calculating the renormalization of the operators $\rm O_4, O_5$  in a similar manner,  one obtains the same result without any mixing complications, ie $\gamma_3 = \gamma_4 = \gamma_5$.
The contribution of each field to the anomalous dimension is the Casimir invariant
of the field's representation times the factor $- \alpha_s (\mu) \log \left[ \mu^2/M_Z^2 \right] / (2 \pi)$.
This result for the renormalization scale dependence agrees with \cite{CataniResum,BauerSchwartz}.
The remaining logs depend on labels which give the phase space configuration 
of the operator.

\subsection{Running the Three Jet Operators}\label{running}

We have matched onto the operators $\rm O_3, O_4, O_5$ at the scale $\mu^2 = M_Z^2$. 
One can run the Wilson coefficients of the three jet
operators using the standard result 
\begin{eqnarray}
\frac{C_i (\mu_2)}{C_i (\mu_1)} = \exp \left[ \int_{\mu_1}^{\mu_2} \frac{d \, \mu}{\mu} \gamma_3(\mu) \right].
\end{eqnarray}
We retain the logs dependent on $n_i$ when solving the RGE as they 
are $\mathcal{O}(1)$.\footnote{We thank A. Manohar for conversations on this point.}
Consider the logarithms dependent on $n_i$.
Expressing the  inner products in terms of the angle between the vectors one finds a divergent log as 
$\theta_{12} \rightarrow 0$,
\begin{eqnarray}
L[(P_1,n_1), (P_2,n_2)]   = \log\left[\frac{2 \,  n_j \cdot P_j  + (1+ \cos \, \theta_{ij})\, \left(n_i \cdot P_i \right)}{2 \, (1- \cos \, \theta_{ij}) \, n_j \cdot P_{j}}  \right]. 
\end{eqnarray}
This limit is not of concern for the $C_F$ phase space logs, as when the gluon becomes collinear with the quark or anti-quark the fermion jets are back to back. The 
$C_A$ phase space logs do diverge in this limit.  

We find that the Wilson coefficients $C_i$, for two stage running, are
\begin{eqnarray}
C_{i}(\mu,n_i) =  \left(\frac{\alpha_s(\mu^2)}{\alpha_s(M_Z^2)} \right)^{\frac{1}{\beta_0} V_1(n_i)} 
\, \left(\frac{\mu^2}{M_Z^2}\right)^{\left(\frac{- 2 \, C_F}{\beta_0} \right) } \,  \left(\frac{\mu^2}{M_Z^2}\right)^{\left(\frac{-C_A}{\beta_0} \right)} C_i(M_Z),
\end{eqnarray}
where
\begin{eqnarray}
V_1(M_Z, n_i) &=& C_F \left( 3 + 2 L[(P_1,n_1), (P_2,n_2)] \right) 
 - \frac{8 \, \pi( C_F + C_A/2)}{\beta_0 \, \alpha_s(M_Z^2)} \\
 &\,& + C_A \left( 1 +   L[(P_1,n_1), (P_3,n_3)] +  L[(P_2,n_2), (P_3,n_3)] -  L[(P_1,n_1), (P_2,n_2)]  \right). \nn
\end{eqnarray}

\section{The Mixing of $\rm O_2$ and $\rm O_3$}\label{mixing}
As one runs the jet operators down to a lower scale, eventually the phase space
scale defining the jet cone is reached. When one runs below this scale,  a third collinear
direction can be resolved in the dijet. Consequently the two and three jet operators mix beginning at the scale $\mu_{\delta}^2$ . 
The following diagrams are defined in the effective theory by taking the limit
before the loop integrals are performed. For the interacting collinear directions $(n_i,n_j)$ this limit is given by
\begin{eqnarray}
\lim_{\phi \rightarrow 0} \mathcal{A} \left( n_i \cdot \bar{n}_j,  n_i \cdot n_j, n_k \cdot n_{i/j} \right) = \lim_{\phi \rightarrow 0} \mathcal{A} \left( 1 + \cos \left(2 \delta + \phi \right),1 - \cos \left(2 \delta + \phi \right) , 1 - \cos \left(\pi - \delta - \phi/2 \right)  \right) \nn
\end{eqnarray}
where  $0 \ll \phi \ll \delta$. This determines the mixing of the two and three jet operators at leading order in $\lambda$ and fixes a $\rm SW$ jet cone in the effective theory.
\begin{figure}[htbp]\label{intout}
\centerline{\scalebox{1.0}{\includegraphics{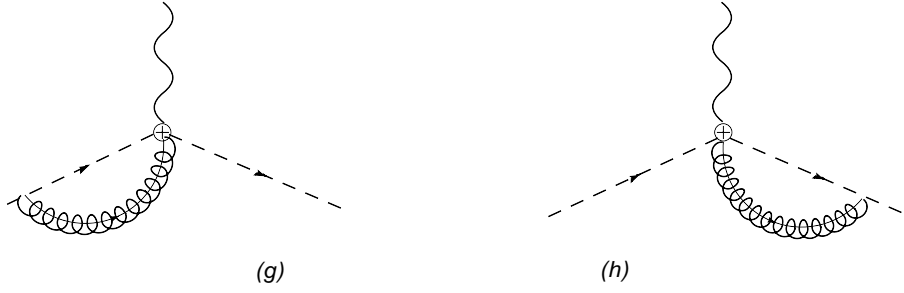}}}
\caption{The mixing of the three and two jet operators.}
\end{figure}
We find  the following mixing for $\rm O_3$ and  $\rm O_2$
\begin{eqnarray}
{\langle {\xi}_{n_1} \, {\bar{\xi}}_{n_2} | {\rm {O}^{\sigma}_3} | 0 \rangle}_{h}(\mu_{\delta}^2) &=&
g_s^2 \, C_F \, 2 \, {\langle {\xi}_{n_1} \, {\bar{\xi}}_{n_2} | {\rm {O}^{\sigma}_2} | 0 \rangle}(\mu_{\delta}^2) \, \int \left( \frac{d \, k}{2 \, \pi} \right)^d \,  \lim_{\phi \rightarrow 0} \, \frac{1}{\bar{n}_3 \cdot k + i \, \epsilon} \, \frac{\bar{n}_1 \cdot P_q + \bar{n}_1 \cdot k }{(P_q + k)^2 + i \, \epsilon} \, \frac{1}{k^2+ i \, \epsilon} \nn \\
&=& \frac{{\langle {\xi}_{n_1} \, {\bar{\xi}}_{n_2} | {\rm {O}^{\sigma}_2} | 0 \rangle}(\mu_{\delta}^2) \, C_F \, g_s^2 }{8 \, \pi^2}
\left( \frac{1}{\epsilon^2} + \frac{1}{\epsilon}  -  \left( \frac{1}{\epsilon} +1 \right)
\log \left[ \frac{-P_{q}^2}{\mu_{\delta}^2} \right] +  \frac{1}{2} \log^2 \left[ \frac{-P_{q}^2}{\mu_{\delta}^2} \right]  \right)  \nn \\
&+&
\frac{{\langle {\xi}_{n_1} \, {\bar{\xi}}_{n_2} | {\rm {O}^{\sigma}_2} | 0 \rangle}(\mu_{\delta}^2) \, C_F \, g_s^2 }{8 \, \pi^2}
\left(2 - \frac{ \pi^2}{12} \right).
\end{eqnarray}
Diagram (g) is given by this result with $P_q \rightarrow P_{\bar{q}}$.  Note that these results agree with the 
form obtained in Eq.(\ref{diagrama}). Implementing the zero bin subtractions in these integrals ensures that the
divergences present are all $\rm UV$ \cite{zerobin}. The mixing is given by the sum of $(g,h)$
\begin{eqnarray}
{\langle {\xi}_{n_1} \, {\bar{\xi}}_{n_2} | {\rm {O}^{\sigma}_3} | 0 \rangle}^{(0)}(\mu_{\delta}^2) 
&=& \frac{{\langle {\xi}_{n_1} \, {\bar{\xi}}_{n_2} | {\rm {O}^{\sigma}_2} | 0 \rangle}(\mu_{\delta}^2) \, C_F \, g_s^2 }{8 \, \pi^2}
\left( \frac{2}{\epsilon^2} + \frac{2}{\epsilon}  -  \left( \frac{1}{\epsilon} +1 \right) \left(
\log \left[ \frac{-P_{q}^2}{\mu_{\delta}^2} \right] + \log \left[ \frac{-P_{\bar{q}}^2}{\mu_{\delta}^2} \right] \right) \right) \nn \\
&+&
\frac{{\langle {\xi}_{n_1} \, {\bar{\xi}}_{n_2} | {\rm {O}^{\sigma}_2} | 0 \rangle}(\mu_{\delta}^2) \, C_F \, g_s^2 }{8 \, \pi^2}
\left(4 - \frac{ \pi^2}{6} + \frac{1}{2} \log^2 \left[ \frac{-P_{\bar{q}}^2}{\mu_{\delta}^2} \right] + \frac{1}{2} \log^2 \left[ \frac{-P_{q}^2}{\mu_{\delta}^2} \right] \right).
\end{eqnarray}
This renormalized matrix element is defined by $ {\langle {\xi}_{n_1} \, {\bar{\xi}}_{n_2} | {\rm {O}^{\sigma}_3} | 0 \rangle}^{(r)} $ being 
free of UV divergences where
\begin{eqnarray}\label{mixing32}
 {\langle {\xi}_{n_1} \, {\bar{\xi}}_{n_2} | {\rm {O}^{\sigma}_3} | 0 \rangle}^{(r)}(\mu_{\delta}^2) &=& \frac{\sqrt{Z_q} \sqrt{Z_{\bar{q}}} \, {\langle {\xi}_{n_1} \, {\bar{\xi}}_{n_2} | {\rm {O}^{\sigma}_3} | 0 \rangle}(\mu_{\delta}^2)}{Z_{3,2}} \nn \\
&=& \frac{{\langle {\xi}_{n_1} \, {\bar{\xi}}_{n_2} | {\rm {O}^{\sigma}_2} | 0 \rangle}(\mu_{\delta}^2) \, C_F \, g_s^2 }{8 \, \pi^2 \, Z_{3,2}}
\left( \frac{2}{\epsilon^2} + \frac{3}{2 \, \epsilon}  -  \left( \frac{1}{\epsilon} +\frac{3}{4} \right) \left(
\log \left[ \frac{-P_{q}^2}{\mu_{\delta}^2} \right] + \log \left[ \frac{-P_{\bar{q}}^2}{\mu_{\delta}^2} \right] \right) \right) \nn \\
&+&
\frac{{\langle {\xi}_{n_1} \, {\bar{\xi}}_{n_2} | {\rm {O}^{\sigma}_2} | 0 \rangle}(\mu_{\delta}^2) \, C_F \, g_s^2 }{8 \, \pi^2 \,  Z_{3,2}}
\left(\frac{9}{2} - \frac{ \pi^2}{6} + \frac{1}{2} \log^2 \left[ \frac{-P_{\bar{q}}^2}{\mu_{\delta}^2} \right] + \frac{1}{2} \log^2 \left[ \frac{-P_{q}^2}{\mu_{\delta}^2} \right] \right).
\end{eqnarray}
The matrix element requires the following counter term
\begin{eqnarray}
Z_{3,2}(\mu_{\delta}^2) =1 + \frac{\alpha_s(\mu_{\delta}^2) \, C_F}{2 \, \pi} \left(\frac{2}{\epsilon^2} + \frac{3}{2 \, \epsilon} - \frac{1}{ \epsilon}
 \left( \log \left[ \frac{-P_{\bar{q}}^2}{\mu_{\delta}^2} \right]
 + \log \left[ \frac{-P_{\bar{q}}^2}{\mu_{\delta}^2} \right] \right)  \right).
\end{eqnarray}
Taking $ P_{q}^2 = P_{\bar{q}}^2 = \delta^2 \, M_Z^2$ we find the off diagonal term of the anomalous dimension matrix,
\begin{eqnarray}
\gamma_{3,2}(\mu_{\delta}) &=& - \frac{3 \, \alpha_s(\mu_{\delta}) \, C_F }{2\,  \pi}.
\end{eqnarray}
For $\rm O_4$ and $\rm O_5$, the mixing of these operators into the two jet operator
vanishes at $\orderalpha$. This occurs due to the operator Dirac structure, demonstrating the utility of 
simplifying the Dirac structure of the three jet operators.

We can now determine the renormalized contribution of the three parton amplitude to the dijet decay rate.
The contribution is the finite real part of Eq.(\ref{mixing32}), given by
\begin{eqnarray}
\langle {\xi}_{n_1} \, {\bar{\xi}}_{n_2} | ({\rm {O}^{\sigma}_3})^{(r)} | 0 \rangle (\mu_{\delta}^2)  =
\langle {\xi}_{n_1} \, {\bar{\xi}}_{n_2} | \rm {O}^{\sigma}_2   | 0 \rangle \,  (\mu_{\delta}^2) \, \hat{C}_3(\mu^2_{\delta})  \, \frac{ \alpha_s(\mu^2_{\delta}) \, C_F }{2\, \pi} \left(\frac{9}{2}  -  \frac{7 \, \pi^2}{6} \right).
\end{eqnarray}

\section{Phase Space Renormalization Group}\label{PSRG} 

To determine $\Gamma(Z \rightarrow J \,  \bar{J})(\mu^2_{\delta})$ we integrate over the two body phase space in $d$ dimensions. The tree level result is
\begin{eqnarray}
 \sigma_0(\epsilon) &=& \int \frac{d \, p_1^3}{(2 \, \pi)^3 \, (2 \, E_1)} \, \frac{d \, p_2^3}{(2 \, \pi)^3 \, (2 \, E_2)}  \sum_{states,pol} (\langle O_2^{\sigma}  \rangle)^2   \nn \\ &=&  \frac{N_C}{32 \, \pi^2}(g_V^2 + g_A^2)[M_Z^{1-2 \,\epsilon} (4 \pi)^{2 \,\epsilon} \frac{2- 2 \,\epsilon}{3 - 2 \,\epsilon} \Omega_{3-2 \, \epsilon} ] \nn \\
&=&   \frac{N_C \, M_z}{12 \, \pi}(g_V^2 + g_A^2) + {\mathcal{O}}(\epsilon),
\end{eqnarray}
so that
\begin{eqnarray}
\Gamma(Z \rightarrow J \,  \bar{J})(\mu^2_{\delta}) = \frac{N_C \, M_z}{12 \, \pi}(g_V^2 + g_A^2) \left(C_2^{two}(\mu^2_{\delta}) +  \hat{C}_3(\mu^2_{\delta}) \, \frac{\alpha_s(\mu^2_{\delta}) \, C_F}{2\, \pi}\left(\frac{9}{2} -  \frac{7 \, \pi^2}{6} \right)\right)^2.
\end{eqnarray}
Reexpanding this expression into fixed order perturbation theory, neglecting non logarithmic ${\mathcal{O}}(\alpha_s(M_Z))$ terms, gives
\begin{eqnarray}
\Gamma(Z \rightarrow J \,  \bar{J})(M_Z^2) = \frac{N_C \, M_z}{12 \, \pi}(g_V^2 + g_A^2) \left( 1 +  \frac{\alpha_s(M_Z^2) \, C_F}{\pi} \left( - 3 \, \log (\delta) - 2 \, \log^2 (\delta) \right) \right).
\end{eqnarray}
Comparing this result to Eq.(\ref{sterW}), we see that we produce the purely collinear contribution $-3 \, \log [\delta]$ correctly, however the double Sudakov log in 
not correctly reproduced.  The reason for this is that only part of the double log is present. We cannot run the collinear degrees of freedom of the dijet operator below the $\mu_{\delta}^2$ scale. 
However, the usoft degrees of freedom run below this scale and are restricted by the $\rm SW$ cuts to lie below the scale $\mu_\beta^2$.  These scales are related by $\mu_\beta^2 \propto \delta^2 \, \mu_{\delta}^2$. To obtain the correct fixed order perturbative expansion for $\Gamma(Z \rightarrow J \,  \bar{J})$,
this restriction of the usoft degrees of freedom must be included.  Including this restriction, we find the correct expression to be given 
by Eq.(\ref{PsRGc2}). 

To determine the correct result, we implement these constraints with 
a one stage running formalism using a Phase Space Renormalization Group ({\rm PSRG}).\footnote{The {\rm PSRG} is inspired by the velocity renormalization group, although some important differences in implementation exist. See \cite{Luke:1999kz} for the 
VRG in NRQCD. The arguments of \cite{Manohar:2000mx,Manohar:2000rz,flemming0306,Pineda01} on the equivalence of one stage and two stage indicate that two stage running could be used to obtain the same results.} Running in this one stage formalism, the collinear degrees of freedom of the dijet run to the scale $\mu_{\delta}^2$;  simultaneously, the usoft degrees of freedom run
to the scale $\mu_\beta^2$.

The usoft matrix element  for the dijet is defined in \cite{SCETJet1,sterman95}. The matrix element depends on the momenta  $k$ of the
usoft degrees of freedom where 
\begin{eqnarray}
 \sum_{X_u} |\langle \, X_u (k) | O_2^{\sigma} | 0 \rangle|^2 \equiv \frac{S(k)}{(Z_2^s(\mu))^2}. 
\end{eqnarray}
We have indicated the required renormalization of the dijet shape function. The usoft radiation  
has momenta $ k = M_Z ( \lambda^2, \lambda^2, \lambda^2)$ where
$\lambda \sim \sqrt{\LQCD/M_Z}$ and the cuts are taken so that $\delta \gg \LQCD/M_Z$.
The cuts being larger than a typical component of an usoft momentum, in the effective theory,
no cut restriction\footnote{However, $|X_u \rangle$ is still not a color singlet.} is placed on the sum over $X_u$. 
We have 
\begin{eqnarray}
S(k)  = \frac{1}{N_C} \, \int \frac{d \, u}{2 \, \pi} e^{i \, k \, u} \langle 0 | \bar{T} \,  [{Y_{n \, d}}^e \, {Y_{\bar{n} \, e}^\dagger}^a] (\frac{u \,n}{2}) \, T \, [{Y_{\bar{n} \, a}}^c \, {Y_{n \, c}^\dagger}^d] (0)| 0 \rangle
\end{eqnarray}
using the notation of  \cite{SCETJet1}. 

The renormalization of the dijet shape function can be determined from $Z_2$.  The 
BPS field redefinition \cite{SCETlukebauer01} is given for all collinear directions $\rm n$ by
\begin{eqnarray}
\xi_n &\rightarrow& Y^{\dagger}_n \, \xi'_n,  \nn \\
\bar{\xi}_n &\rightarrow& \xi'_n Y_n, \\
A_n &\rightarrow& Y^{\dagger}_n \, A'_n Y_n,
\end{eqnarray}
where the path-ordered Wilson line of usoft gluons in the
$n$ direction is
\begin{eqnarray}
Y_n (z) = P \, {\rm exp} \left[ i \, g \, \int_0^\infty \, d s \, n \cdot A_u (n s + z) \right].
\end{eqnarray}
This field redefinition removes all couplings in the Lagrangian of 
the redefined usoft and collinear fields $( \xi'_n, \xi'_n, A'_n)$. This allows the matrix element to be factorized
\begin{eqnarray}
\sum_{final \, states} |\langle J_n \, J_{\bar{n}} \, X_u |  {\bar {\chi}}_{n_q} \, \Gamma^{\sigma} \,  {\chi}_{n_{\bar{q}}} (0) \, \epsilon_{\sigma} \, |0 \rangle|^2 &=& 
\sum_{J_n \, J_{\bar{n}} } |\langle J_n \, J_{\bar{n}} | {\bar {\chi'}}_{n_q} \, \Gamma^{\sigma} \,  {\chi'}_{n_{\bar{q}}} (0)  \, \epsilon_{\sigma} \, |0 \rangle|^2 \, \nn \\
&\,& \times \sum_{X_u} |\langle X_u | T \, [{Y_{\bar{n} }} \, {Y_{n \, }^\dagger}] (0) |0 \rangle|^2.
\end{eqnarray}
The renormalization of the initial matrix element must be reproduced by the 
product of the renormalized matrix elements after the field redefinition. Thus
$Z_2 = Z_2^c \, Z_2^u$, where $Z_2^c$ is the renormalization for the matrix
element of only collinear degrees of freedom, while $Z_2^s$ gives the renormalization 
for the usoft matrix element. These renormalization factors are 
\begin{eqnarray}
Z_{2}^c(\mu_c)&=& 1 + \frac{\alpha_s(\mu_c^2) \, C_F}{2 \, \pi} ( \frac{2}{\epsilon^2} + \frac{3}{2 \,\epsilon} - \frac{1}{\epsilon}  \ln{[\frac{p_q^2 \,  p_{\bar{q}}^2}{\mu_c^4}]} ) + \orderalphasqr,
\end{eqnarray}
and 
\begin{eqnarray}
Z_{2}^s(\mu_s)&=& 1 + \frac{\alpha_s(\mu_s^2) \, C_F}{2 \, \pi} ( -\frac{1}{\epsilon^2}  + \frac{1}{\epsilon}  \ln{[\frac{n \cdot p_q \,  \bar{n} \cdot p_{\bar{q}}}{\mu_s^2}]} ) + \orderalphasqr.
\end{eqnarray}
Note that the convolution of the variable $k$ between the jet and usoft matrix element is neglected as we are considering the total decay rate.
The anomalous dimensions of the factorized matrix elements are given by
\begin{eqnarray}
\gamma_{2}^c(\mu_c)&=&  - \frac{\alpha_s(\mu_c^2) \, C_F}{ \pi} (- \ln{[\frac{p_q^2 \,  p_{\bar{q}}^2}{\mu_c^4}]} + \frac{3}{2} ) +  \orderalphasqr,
\end{eqnarray}
\begin{eqnarray}
\gamma_{2}^s(\mu_s)&=&  - \frac{\alpha_s(\mu_s^2) \, C_F}{ \pi} (  \ln{[\frac{n \cdot p_q \,  \bar{n} \cdot p_{\bar{q}}}{\mu_s^2}]} ) +  \orderalphasqr.
\end{eqnarray}
The phase space scales that we will run the usoft and collinear factorized matrix elements to are related by $\mu^2_{\beta} = \mathcal{B}^{\delta -1} \, \mu^4_{\delta} / (M_Z^2)$. We relate both $\mu_s$ and $\mu_c$ to a 
phase space parameter $\phi$ that runs from $1$ to $\delta$ using
\begin{eqnarray}
\mu_c^2 &=&\phi^2 \, M_Z^2, \nn \\
\mu_s^2 &=&  \phi^4 \, M_Z^2 \, \mathcal{B}^{(\phi -1)}.
\end{eqnarray}
We can express the running of $\rm C_2$ in the {\rm PSRG} as
\begin{eqnarray}
\phi \, \frac{d}{d \, \phi} C_2(\phi) = (\gamma_{2}^c(\phi) + \left(2 + \frac{\phi}{2} \, \log ( \mathcal{B}) \right) \, \gamma_{2}^s(\phi))C_2(\phi).
\end{eqnarray}
We take the invariants to be $P_q^2, P_{\bar{q}}^2 = \delta^2 M_Z^2$ and $n \cdot p_q  \, \bar{n} \cdot p_{\bar{q}}  = f({\mathcal{B}}) \, \delta^4 \, M_Z^2 $ giving
\begin{eqnarray}
\phi \, \frac{d}{d \, \phi} C_2(\phi) &=& - \frac{\alpha_s(M_Z \, \phi) \, C_F}{\pi} \left(\frac{3}{2} - 2 \, \log \left[\frac{\delta^2}{\phi^2} \right]  \right) C_2(\phi) \\
&-&\frac{\alpha_s(M_Z \, \phi) \, C_F}{\pi}  \left(2+ \frac{\phi}{2} \, \log \left({\mathcal{B}} \right) \right)
\left(  \log \left( \frac{ f({\mathcal{B}}) \delta^4}{\phi^4 \, {\mathcal{B}}^{\phi - 1}} \right)
\right)  C_2(\phi).
\end{eqnarray}
The particular functional dependence on ${\mathcal{B}}$ in the logarithm is fixed by
matching onto the full theory. Note that we must refer to QCD with the SW jet definition at the collinear scale as we matched the amplitude in QCD onto our SCET operator basis. As this matching 
was performed before the phase space integrals are performed we must refer to the full theory to establish the jet definition at the collinear scale.
As in SW, we expand
in the dependence on ${\mathcal{B}}$ and $\delta$ and neglect the dependence on these parameters that does not contribute to large logarithms. 

This RGE is solved using standard methods to give
\begin{eqnarray}\label{PsRGc2}
C_2^{SW}(\mu^2_{\beta} , \mu^2_{\delta} ) =   C_{2}(M_Z) \,  \left(\frac{\alpha_s(\delta^2 \, M_Z^2)}{\alpha_s(M_Z^2)} \right)^{\frac{C_F}{\beta_0}\left(3 + \frac{16 \, \pi}{\beta_0 \, \alpha_s(\delta^2 \, M_Z^2  f({\mathcal{B}})}) \right)}  \, \left(\delta^2\right)^{\left(\frac{4 \, C_F}{\beta_0} \right)}. 
\end{eqnarray} 
The contributions from the running of the three jet operators contribute at $\orderalphasqr$. We choose to retain two stage running for the three 
jet contribution to the Sudakov resumed dijet decay rate. 

 \section{Two Jet Decay Rate}\label{results}

Using  $C_2^{SW}(\mu^2_{\beta} , \mu^2_{\delta} ) $ we obtain the total decay rate 
\begin{eqnarray}
\Gamma(Z \rightarrow J \,  \bar{J})(\mu^2_{\beta} , \mu^2_{\delta} ) = \int \, \sum_{states} |\langle O_2^\sigma \rangle|^2(\mu^2_{\delta})  \left(C_2^{SW}(\mu^2_{\beta} , \mu^2_{\delta} )  +  \frac{\alpha_s(\mu^2_s) \, \hat{C}_3 \, C_F}{2\, \pi}\left(\frac{9}{2} -  \frac{7 \, \pi^2}{6} \right)\right)^2,
\end{eqnarray}
where we have indicated the dependence on the perturbative expansion of $|\langle O_2^\sigma \rangle|^2$. The contributions to this expression at $\orderalpha$ come from the running of $\rm O_2$ in the PSRG and the mixing of $\rm O_3$ and $\rm O_2$ beginning at the collinear scale. 
This is sufficient to resum the double Sudakov logarithm and the class of sub leading logs we are interested in. The running of $\rm O_3$ contributes logarithms that first contribute at $\orderalphasqr$. To extend the resummation consistently to sub-leading logarithms, the running of $\rm O_3$
must be reexpressed in the PSRG. This is beyond the scope of this paper.

The renormalized perturbative expansion of $|\langle O_2^\sigma \rangle|^2$ is $\rm IR$ finite,  as the $\rm SW$ jet definition was constructed to satisfy the KLN theorem \cite{LeeNauenberg, kinoshita}.
In our effective theory construction,  the KLN theorem for the $\rm SW$ states ensures that the dijet partial cross section is free of explicit $\rm IR$ singularities, as the $\rm IR$ of the 
full theory is reproduced in the effective theory.

The expansion of the operator gives the collinear gluon emission contained within the dijet cones.  However, the fixing of the SW definition within the effective theory removes gluons that are excluded by the $\rm SW$ jet definition from the jet cones.
This results in constant terms due to the degrees of freedom removed by the jet definition 
that have to be accounted for by matching in the effective theory.
We introduce the phase space Wilson coefficient $\mathcal{S}_2(\delta^2 \, M_Z)$ to account for this further matching onto the jet definition in the full theory. As a result, we have
\begin{eqnarray}
C_2^{SW}(\mu^2_{\beta} , \mu^2_{\delta} ) =   C_{2}(M_Z) \,  S_{2}(\delta^2 \, M_Z)  \left(\frac{\alpha_s(\delta^2 \, M_Z^2)}{\alpha_s(M_Z^2)} \right)^{\frac{C_F}{\beta_0}\left(3 + \frac{16 \, \pi}{\beta_0 \, \alpha_s(\delta^2 \, M_Z^2 f({\mathcal{B}})}) \right)}  \, \left(\delta^2\right)^{\left(\frac{4 \, C_F}{\beta_0} \right)}. 
\end{eqnarray}
We determine $\mathcal{S}_2(\delta^2 \, M_Z) $ by calculating the SW dijet decay rate in full QCD, and 
in SCET with the PSRG. We match onto the SW jet definition in the full theory to determine the perturbative expansion of  $\mathcal{S}_2(\delta^2 \, M_Z) $.

Computing the $\rm SCET$ graphs in dimensional regularization we find the following $\orderalpha$ corrections
to the matrix element \footnote{See \cite{zerobin} for details on these effective theory calculations and the requirement of zero bin subtractions. Note that as we are integrating over the 
full phase space. To avoid double counting we find the zero bin subtracted contribution to be $R_F+ R_B - R_A - 1/2 R_C - 1/2 R_E $ in the notation of \cite{zerobin} .}
\begin{eqnarray}
\int \, \sum_{states} |\langle O_2^\sigma \rangle|^2(\mu^2_{\delta}) = \sigma_0(\epsilon) \left(1 + 2  \, \frac{Z_{\xi}^4(\mu^2_{\delta})}{Z_2^2 (\mu^2_{\delta})}  \, \int_0^1 \, d \, x_1 \, \int_{(1- x_1)}^1 \, d \, x_2 \, S(x_1, x_2,\mu^2_{\delta}) \, A(x_1,x_2) \right)      
\end{eqnarray}
with 
\begin{eqnarray}
S(x_1, x_2,\mu^2_{\delta}) &=& \frac{\alpha_s(\mu^2_{\delta}) \, C_F}{2 \, \pi} \, \frac{1}{\Gamma(1- \epsilon) \, (1- x_1)^\epsilon \,  (1- x_2)^\epsilon \, (x_1+ x_2 -1)^\epsilon}, \nn \\
A(x_1, x_2) &=& \frac{1+ x_2^2}{(1-x_1)(1-x_2)} - \frac{\epsilon \, (1- x_2)}{(1- x_1)} - \frac{1}{(1-x_1)\, (1-x_2)} - \frac{1 - \epsilon}{2 \, (1- x_1)}.
\end{eqnarray}
Integrating these expressions we find 
\begin{eqnarray}
\int \, \sum_{states} |\langle O_2^\sigma \rangle|^2(\mu^2_{\delta}) = \sigma_0(\epsilon) \left(1 +  \frac{\alpha_s(\mu^2_{\delta}) \, C_F}{2 \, \pi} \left(\frac{23}{2} - \frac{7 \, \pi^2}{6} \right) \right).
\end{eqnarray} 
To check this result we re-expand the Sudakov factors out into fixed order perturbation theory  in terms of $\alpha_s(M_Z)$, we find
\begin{eqnarray}\label{stermanW}
\Gamma(Z \rightarrow J \bar{J} )(M_z) &=& \frac{N_C \, M_z (g_V^2 + g_A^2)}{12 \, \pi}\left(1 + \frac{\alpha_s(M_z) \, C_F }{\pi} \left[- 3 \ln \delta - 4 \ln (f({\mathcal{B}}) \,  \delta) \ln (\delta) \right] \right) \\
 &+& \frac{N_C \, M_z (g_V^2 + g_A^2)}{12 \, \pi}\left(1 + \frac{\alpha_s(M_z) \, C_F }{\pi} \left[ \frac{57}{4} - \frac{7 \, \pi^2}{3} \right]  + 2 \,  \alpha_s(M_z) \, s_1  \right) + \orderalphasqr .\nn
\end{eqnarray} 
We match onto the $\rm SW$ dijet decay rate to obtain $f({\mathcal{B}}) \,  \delta \equiv 2 \,  \beta $.
The phase space Wilson coefficient $\mathcal{S}_2(\delta^2 \, M_Z)$ has the
perturbative expansion
\begin{eqnarray}
\mathcal{S}_2(M_Z) = 1 + \alpha_s (M_z ) \, s_1  + \orderalphasqr,
\end{eqnarray} 
where 
\begin{eqnarray}
s_1 = C_F \left( \pi - \frac{47}{8 \, \pi} \right).
\end{eqnarray} 
This is the main result of our paper. The Sudakov resumed dijet decay rate in $\rm SCET$ is given  by
\begin{eqnarray}
\Gamma(Z \rightarrow J \,  \bar{J})(\mu^2_{\beta} , \mu^2_{\delta} ) &=& \sigma_0 \, \left(1 +  \frac{\alpha_s(\mu^2_{\delta}) \, C_F}{2 \, \pi} \left(\frac{23}{2} - \frac{7 \, \pi^2}{6} \right) \right) \\
&\, & \times \left(C_2^{SW}(\mu^2_{\beta} , \mu^2_{\delta} )  +  \frac{\alpha_s(\mu^2_s) \, \hat{C}_3 \,  C_F}{2\, \pi}\left(\frac{9}{2} -  \frac{7 \, \pi^2}{6} \right)\right)^2.\nn
\end{eqnarray}

Comparing to the literature, we note that a leading log resummation was determined in \cite{mukhisterman} in full QCD for the SW jet definition.  The exponentiation of the double Sudakov logarithm agrees in both formalisms. Comparing the results, we consider the resummation in  SCET with the PSRG to be conceptually clearer and easier to extend to non leading logarithms.

\section{Conclusions}
We have determined the RGE improved decay rate $\Gamma(Z \rightarrow J \, \bar{J})$
using $\rm SCET$.  Recalling the $\rm SW$ jet definition, we can now  
state the corresponding definition of the dijet final states in our effective theory.

 The dijet decay rate comes from the direct matching of $\rm O_2^{\sigma} $ onto $\rm QCD$
and from the mixing of the three jet operators at the 
scale defining the angular cut $\mu^2_{\delta}$. These contributions correspond to the energetic quarks of SW1 and the gluons of collinear energy emitted inside 
the jets, SW3.

 The nonperturbative corrections given by matrix elements of the ultrasoft
final state gluons are restricted by the scale $\mu^2_{\beta}$.
These final state gluons correspond to the emission of gluons unrestricted in direction 
with low energy. The ultrasoft matrix element with this restriction corresponds to the state SW2. 

Treating the decay rate in $\rm SCET$ allowed the dijet decay amplitude 
to be defined in terms of $\rm SCET$ jet fields. The EFT formalism 
allowed this definition to take place without large logarithms.  The 
fraction of three parton events contributing to the dijet rate in the $\rm SW$ jet definition,
corresponds to the three jet operators mixing into the dijet operators
at the scale $\mu^2_{\delta}$, and the perturbative expansion of  $\rm O_2^{\sigma} $ at the scale $\mu^2_{\delta}$. The general version of this result is that $n+i$ jet operators mix into 
$n$ jet cross sections and decay rates at order $\alpha_s^i(\mu^2_\delta)$.

To obtain the resummed dijet decay rate, we introduced a one stage running formalism for the 
dijet operator. This allowed us to run to the phase space cut scales defining 
the dijet corresponding to the SW jet definition. We also introduced phase space Wilson coefficients. These are 
required to fix the SW jet definition onto the jet operators in SCET.

This approach establishes a jet definition in $\rm SCET$ and resums the 
large phase space logarithms that come about as the result of the jet definition.
This allows a program of systematically improving the perturbative behavior of jet
observables to be carried out in SCET.

\section{Acknowledgments}
\label{acknowledgments}
We acknowledge helpful and prescient conversations with Aneesh Manohar on this subject,
and thank him for encouragement. We also acknowledge conversations with
Christian W. Bauer, B. Grinstein, I.W. Stewart and M. Luke on this material. We further thank Aneesh Manohar, Alex Williamson and 
M. Luke for comments
on the manuscript.

As this paper was approaching completion the work \cite{BauerSchwartz2}, which is a more detailed accounting of the work
reported on in  \cite{BauerSchwartz} appeared. In \cite{BauerSchwartz2} the calculation of the renormalization of 
three jet operators equivalent to the three jet operators examined in this paper, Eq.(\ref{threejrenorm}), was reported. We find that \cite{BauerSchwartz2},
which was not calculated in the background field method, is inconsistent in its inclusion of the wavefunction renormalization of the gluon
while not including the QCD coupling constant renormalization. Once \cite{BauerSchwartz,BauerSchwartz2} is corrected for this inconsistency, their results agree with ours.

\appendix

\section{Feynman rules of the 3-Jet Operators}
For completeness, we state the one and two gluon Feynman rules of the operators $\rm O_3, O_4, O_5$  to ${\mathcal{O}}(\lambda^0 \, M_Z^2)$
in this Appendix, we utilize these rules to renormalize the operators in the Section \ref{renorm}. 
The following Feynman rules are for emitted collinear gluons. 

The one gluon Feynman rules where the gluon is emitted in the direction
$n_3$ requires the collinear gluon field strength to emit a gluon.
Here $P_g,P_q,P_{\bar{q}}$ are outgoing collinear momentum:
\begin{eqnarray}
\langle \bar{\xi}_{n_1} \,  \xi_{n_2} \, A^{\mu}_{n_3} |  O_3^{\sigma}| 0 \rangle &=&  \bar{\xi}_{n_1} \,
2 \, g_s \, T^a \, \left( \frac{ n_1^{\mu}}{(n_1 \cdot n_3) (\bar{n}_3 \cdot P_g)} -  \frac{n_2^{\mu}}{(n_2 \cdot n_3) (\bar{n}_3 \cdot P_g)} \right) \Gamma^{\sigma} \, \xi_{n_2} ,
\nn \\
\langle \bar{\xi}_{n_1} \,  \xi_{n_2} \,  A^{\mu}_{n_3} |  O_4^{\sigma}| 0 \rangle &=&  \bar{\xi}_{n_1} \, g_s \, T^a \,\left( \frac{1}{(n_1 \cdot n_3) (\bar{n}_1 \cdot P_q)} -  \frac{1}{(n_2 \cdot n_3) (\bar{n}_2 \cdot P_{\bar{q}})} \right)
(n_3^{\sigma} \, \Gamma^{\mu} - n_3 \cdot \Gamma \, g^{\sigma \, \mu}) \, \xi_{n_2} ,\nn \\
 \langle \bar{\xi}_{n_1} \,  \xi_{n_2} \,  A^{\mu}_{n_3} |  O_5^{\sigma}| 0 \rangle &=&  \bar{\xi}_{n_1} \,  g_s \, T_a \, \left( \frac{1}{(n_1 \cdot n_3) (\bar{n}_1 \cdot P_q)} -  \frac{1}{(n_2 \cdot n_3) (\bar{n}_2 \cdot P_{\bar{q}})} \right)
 i \, \epsilon^{\alpha \, \beta \, \sigma \, \eta} (n_3)_{\beta} \, g_{\alpha \, \mu} \, \gamma_5 \Gamma_{\eta} \, \xi_{n_2} \nn.
\end{eqnarray}

The two gluon Feynman rules with $P_g^1$ and $P_g^2$
the two outgoing collinear gluon momenta, associated with
$A^{\mu}_a$ and  $ A^{\nu}_b$ are
\begin{eqnarray}
\langle \bar{\xi}_{n_1} \,  \xi_{n_2} \, A^{\mu,a}_{n_3}  \, A^{\nu,b}_{n_3}  | O_3^{\sigma} | 0 \rangle &=&  \bar{\xi}_{n_1} \, \frac{ 2 \, g_s^2 \, (T^a\, T^b)}{(\bar{n}_3 \cdot P_g^1) \, (\bar{n}_3 \cdot P_g^2)}
\left[ \frac{{\bar{n}_3}^\mu \,  n_1^\nu -n_1^\mu \,  {\bar{n}_3}^\nu }{n_1 \cdot n_3} - \frac{{\bar{n}_3}^\mu \,  n_2^\nu -n_2^\mu \, {\bar{n}_3}^\nu }{n_2 \cdot n_3}\right] \Gamma^{\sigma} \, \xi_{n_2}   \nn \\
&+&  \bar{\xi}_{n_1} \,  \frac{ 2 \, g_s^2 \,  (T^b\, T^a)}{(\bar{n}_3 \cdot P_g^1) \, (\bar{n}_3 \cdot P_g^2)}
\left[ \frac{{\bar{n}_3}^\nu \,  n_1^\mu -n_1^\nu \,  {\bar{n}_3}^\mu }{n_1 \cdot n_3} - \frac{{\bar{n}_3}^\nu \,  n_2^\mu -n_2^\nu \,  {\bar{n}_3}^\mu }{n_2 \cdot n_3}\right] \Gamma^{\sigma} \, \xi_{n_2}  \nn \\
&+&  \bar{\xi}_{n_1} \,  \frac{ 4 \, g_s^2 \,  f^{a\, b \, c} \, T_c \, (n_2^\mu \,  n_1^\nu -n_1^\mu \,  n_2^\nu )}{(\bar{n}_3 \cdot (P_g^1+ P_g^1))^2 \, (n_1 \cdot n_3) \, (n_2 \cdot n_3) } \Gamma^{\sigma} \, \xi_{n_2} 
\end{eqnarray}
\begin{eqnarray}
\langle \bar{\xi}_{n_1} \,  \xi_{n_2} \, A^{\mu,a}_{n_3}  \, A^{\nu,b}_{n_3}  | O_4^{\sigma} | 0 \rangle &=&  \bar{\xi}_{n_1} \, 
\frac{g_s^2 \, i \, f^{a \, b \,c} \, T_c \, {\bar{n}}_3^\mu}{\bar{n}_3 \cdot P_g^1}
\, \left( \frac{\left(n_3^\sigma \, \Gamma^\nu - n_3 \cdot \Gamma \, g^{\sigma \, \nu} \right)}{(n_1 \cdot n_3) (\bar{n}_1 \cdot P_q)}
 -  \frac{\left(n_3^\sigma \, \Gamma^\nu - n_3 \cdot \Gamma \, g^{\sigma \, \nu} \right)}{(n_2 \cdot n_3) (\bar{n}_2 \cdot P_{\bar{q}})} \right) \, \xi_{n_2}  \nn \\
&-&  \bar{\xi}_{n_1} \, 
\frac{ g_s^2 \, i \, f^{a \, b \,c} \, T_c \, {\bar{n}}_3^\nu}{\bar{n}_3 \cdot P_g^2}
\, \left( \frac{\left(n_3^\sigma \, \Gamma^\mu - n_3 \cdot \Gamma \, g^{\sigma \, \mu} \right)}{(n_1 \cdot n_3) (\bar{n}_1 \cdot P_q)}
 -  \frac{\left(n_3^\sigma \, \Gamma^\mu - n_3 \cdot \Gamma \, g^{\sigma \, \mu} \right)}{(n_2 \cdot n_3) (\bar{n}_2 \cdot P_{\bar{q}})} \right) \, \xi_{n_2}  \nn \\
&+&   \bar{\xi}_{n_1} \,  \frac{ 2 \, g_s^2 \,   f^{a\, b \, c} \, T_c }{(\bar{n}_3 \cdot (P_g^1+ P_g^2)} \left( \frac{ \Gamma^{\mu} \, g^{\sigma \, \nu} -  \Gamma^{\nu} \, g^{\sigma \, \mu}}{(n_1 \cdot n_3) \, (n_1 \cdot P_q)} -
\frac{ \Gamma^{\mu} \, g^{\sigma \, \nu} -  \Gamma^{\nu} \, g^{\sigma \, \mu}}{(n_2 \cdot n_3) \, (n_2 \cdot P_{\bar{q}})} \right) \, \xi_{n_2} 
\end{eqnarray}

\begin{eqnarray}
\langle \bar{\xi}_{n_1} \,  \xi_{n_2} \, A^{\mu,a}_{n_3}  \, A^{\nu,b}_{n_3}  | O_5^{\sigma} | 0 \rangle &=&  \bar{\xi}_{n_1} \, 
g_s^2 \, i \, f^{a \, b \,c} \, T_c \, \left( i \, \epsilon^{\alpha \, \beta \, \sigma \, \eta} \, (n_3)_\beta \, \gamma_5 \, \Gamma_\eta \right) \,
\left( \frac{g_{\alpha \, \nu} \, n_3^\mu}{\bar{n}_3 \cdot P_g^1} - \frac{g_{\alpha \, \mu} \, n_3^\nu}{\bar{n}_3 \cdot P_g^2}
\right)  \\
& \times& \left( \frac{1}{(n_1 \cdot n_3) (\bar{n}_1 \cdot P_q)} -  \frac{1}{(n_2 \cdot n_3) (\bar{n}_2 \cdot P_{\bar{q}})} \right) \, \xi_{n_2}  \nn \\
&+& \bar{\xi}_{n_1} \, \frac{2 \, g_s^2 \, f^{a \, b \,c} \, T_c}{ ( \bar{n}_3 \cdot (P_g^1+ P_g^2))}
\, \left( \frac{\left( i \, \epsilon^{\mu \, \nu \, \sigma \, \eta} \, \gamma_5 \, \Gamma_\eta \right)}{(n_1 \cdot n_3)  \,  (n_1 \cdot P_q)}  
- \frac{\left( i \, \epsilon^{\mu \, \nu \, \sigma \, \eta} \, \gamma_5 \, \Gamma_\eta \right)}{(n_2 \cdot n_3) \,  (n_2 \cdot P_{\bar{q}})} \right) \, \xi_{n_2}  \nn
\end{eqnarray}


\end{document}